# MotifMark: Finding Regulatory Motifs in DNA Sequences*

Hamid Reza Hassanzadeh, *Student Member, IEEE*, Pushkar Kolhe, Charles L. Isbell, and May D. Wang, *Senior Member, IEEE*

*Abstract*— The interaction between proteins and DNA is a key driving force in a significant number of biological processes such as transcriptional regulation, repair, recombination, splicing, and DNA modification. The identification of DNA-binding sites and the specificity of target proteins in binding to these regions are two important steps in understanding the mechanisms of these biological activities. A number of high-throughput technologies have recently emerged that try to quantify the affinity between proteins and DNA motifs. Despite their success, these technologies have their own limitations and fall short in precise characterization of motifs, and as a result, require further downstream analysis to extract useful and interpretable information from a haystack of noisy and inaccurate data. Here we propose MotifMark, a new algorithm based on graph theory and machine learning, that can find binding sites on candidate probes and rank their specificity in regard to the underlying transcription factor. We developed a pipeline to analyze experimental data derived from compact universal protein binding microarrays and benchmarked it against two of the most accurate motif search methods. Our results indicate that MotifMark can be a viable alternative technique for prediction of motif from protein binding microarrays and possibly other related high-throughput techniques.

## I. INTRODUCTION

DNA binding proteins are key components of different cellular processes including transcription, translation, repair, and replication machinery. Studies have shown that an estimated 6%-7% of the genome of eukaryotic organisms encode such proteins [1]. One of the important protein-DNA interactions that are vital for expression of genes in the cells is the interaction between transcription factors (TFs) and their corresponding binding sites. Through these sequence-specific interactions, numerous spatial and temporal activities in biological pathways are coordinated and as such finding these sites has a significant clinical value. With advances in high-throughput technologies in the past decade several in-vivo and in-vitro techniques have been invented and upgraded to address this important and yet challenging task. Unfortunately, none of these methods are able to generate results that are directly interpretable by biologists, but instead each generates a large volume of noisy, erroneous and low-resolution measurements for tens of thousands of sequence probes. As a result, the outcome of such experiments needs to be processed through downstream analysis pipelines to elicit useful information. In this study, we use data from Protein Binding Microarray (PBM) experiments to evaluate our proposed method. Protein Binding Microarrays (PBM) [2] and subsequently universal PBM were among the most notable large scale in-vitro technologies that were able to characterize sequence specificities of DNA-protein interactions. In a PBM experiment, TF is allowed to bind to double stranded DNA segments after which the protein-bound array is washed gently and stained with a primary antibody specific to its epitope. Compact universal PBM, on the other hand, uses a compact representation of $k$-mers in longer probes (often 60bp long), and therefore, the experiment can be performed in a cost- and space-effective way. The downside is that now the measured intensity for each probe is affected by all $k$-mers in it. This calls for an accurate motif search pipeline that can characterize the DNA binding sites from these low resolution experiment data.

To date, numerous computational tools have been developed to characterize DNA binding sites from high-throughput techniques (see [3] for a comprehensive assessment of some of the most popular approaches). A majority of these techniques are designed based on position weight matrices (PWM) with different learning algorithms, or at least inspired by them (see [4] for a recent assessment of PWM-based tools.) These methods often work well in practice and their outcomes are interpretable to researchers. Of particular interest to us are RankMotif++ [5] and KmerHMM [6], two prominent computational pipelines for prediction of motifs, that have been shown to produce state-of-the-art performances on PBM experiment data [6]. RankMotif++ adopts a probabilistic approach to model the relative binding preferences between probe pairs and the target TF. KmerHMM, on the other hand, models the motifs as hidden Markov models (HMMs) where each HMM state emits some nucleotide or a gap (for insertions). KmerHMM has been shown to outperform other classical computational approaches on several test benches or generate competitive results, otherwise. The success of KmerHMM, can be attributed to two distinguished features of HMM. First, it reckons with the dependencies between neighboring positions, and second, it can capture the dynamics of multiple motifs through different state paths. Two important shortcomings in most motif prediction tools are, first, they

*This work was supported by the grants from National Institutes of Health (NCI Transformative R01 CA163256, and National Center for Advancing Translational Sciences UL1TR000454), Microsoft Research and Hewlett Packard. The content is solely the responsibility of the authors and does not necessarily represent the official views of the National Institutes of Health.

H. R. Hassanzadeh is with the Department of Computational Science and Engineering, Georgia Institute of Technology, Atlanta, GA 30332 USA. (email: hassanzadeh@gatech.edu).
C. L. Isbell is with the College of Computing, Georgia Institute of Technology, Atlanta, GA 30332 USA. (email: isbell@cc.gatech.edu)
P. Kolhe is with the College of Computing, Georgia Institute of Technology, Atlanta, GA 30332 USA. (email: pushkar@cc.gatech.edu)
M. D. Wang is with the Department of Biomedical Engineering, Georgia Institute of Technology and Emory University and the School of Electrical and Computer Engineering, Georgia Institute of Technology, Atlanta, GA 30332 USA (corresponding author, phone: 404-385-2954; e-mail: maywang@bme.gatech.edu).

don't take into account the fact that PBM assays cannot distinguish the strand on which the motifs are located as PBM deals with dsDNA, and second, they either fail to capture the non-linear relations between different motif base positions, or otherwise, consider simplistic assumptions about them, such as the first order Markovian dependence, as is the case for KmerHMM. The first deficiency can derail the training process towards a local optimum that is biologically irrelevant, while the second can lead to degenerate models, as a result of which competing sequences may not be ranked correctly. More recently, techniques based on deep learning [7] have been introduced to address the above challenges. However, these methods are still in their primitive stages and lack interpretability. In this article, we propose a new computational approach that addresses these shortcomings and demonstrate its promise through multiple comparisons.

## II. MATERIALS AND METHODS

In this study, we propose a new mathematical model that addresses the aforementioned issues through proper employment of graph theory and machine learning.

Figure 1 depicts the computational framework that we developed for our model. For each PBM experiment the data contains a set of probes that correspond with measured signal intensities, $\{I_1, I_2, \ldots, I_N\}$. As a first step, the normalized intensity values are recorded for $k$-mers (and their reverse complement) encoded in experiment probes. Next, we sort the probes based on their measured intensities and generate a set, $M$, comprising the top performing ones. Then, we seek to choose the smallest subset, $T$, of $k$-mers encoded in $M$, such that for every probe in $M$, there exists a $k$-mer in $T$ that either itself or its reverse complement is encoded in that probe. To find the minimum set, $T$, the brute-force approach would be to evaluate all possible subsets of k-mers (requires $\Omega(|M|2^{4^k})$ $operations$) which is computationally infeasible.

Here, we solve a slightly different problem. Instead of trying to find the minimum set we solve a graph matching problem that gives a approximate minimal set. To do that, we

Algorithm 1: Finding a candidate set of positive k-mers

**Data:** A bipartite graph comprising a probe set, $P$, and a $k$-mer set, $T$
**Result:** $S$, the set of positive $k$-mer
$S = \emptyset$;
**while** *not all nodes in P covered* **do**
  Select node $t \in T$ that has the highest degree;
  $S = S \cup \{t\}$;
  Remove all elements in $P$ that are adjacent to $t$;
  Remove all elements in $T$ that are not incident with any edge;
**end**

reformulate the problem setting as a bipartite graph with nodes in its two components representing probes in $M$ and the $k$-mers in $T$, respectively. Moreover, we add an edge between two vertices $p_i$ and $t_j$ iff $t_j \subset_{seq} p_i$ (where $t \subset_{seq} p$ if p contains either t or its reverse complement). Our objective is to pick a subset of edges such that a reasonably small (not necessarily the smallest) number of nodes in the $k$-mer component are covered subject to the condition that all elements of the probe component be covered. Algorithm 1 illustrates an iterative greedy solution to the new problem with polynomial time complexity. In summary, at each step, the $k$-mer which is incident with more probes is selected. This is because probes that are strongly bound to the TF are more likely to share similar $k$-mers. Once selected, we remove all the probes that are covered by this $k$-mer. Note that, at any stage but the first, some orphaned $k$-mers may have remained. These are the ones which have been dominated by more pronounced $k$-mers and are not likely to be valid binding sites.

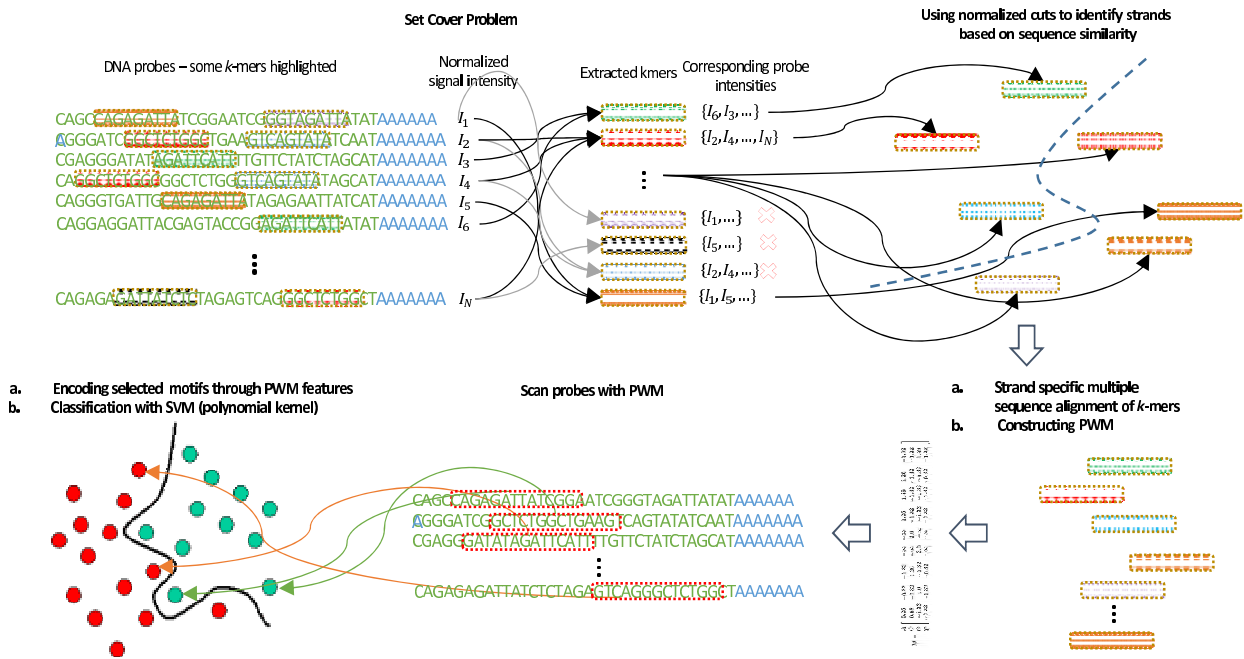

Figure 1: Illustration of major steps in MotifMark.

Once the positive $k$-mer set is constructed we need to align them using multiple sequence alignment techniques. Note, however, that at this stage the strand corresponding to the selected $k$-mers is not known. Aligning the sequences which are not located on the same strand misleads the training process. This is especially the case for non-palindromic motifs. Hence, we used a heuristic approach to divide the set, $T$, into strand specific subsets. To do that, we constructed a complete weighted graph with nodes representing the selected $k$-mers and edge weights being the negated edit distance between the corresponding $k$-mers. The objective now is to cut the graph into two pieces such that the sum of weights on the cut-set is minimized, and hence the name *minimum cut*. There are efficient algorithms that solve this problem. However, the optimal solution to this problem is not frequently appropriate in practice [8] as the minimum cut criterion tends to divide the graph into a small cluster with very few nodes and a large one containing the rest of the nodes. Therefore, instead, we adopted the strategy proposed by Shi et. al. [9] which normalizes the cost function according to the cluster sizes and hence the name, normalized cuts. They showed that this problem is an NP-complete one and offered a greedy approximate solution for it that has been successfully tried by a significant number of studies.

Upon dichotomization of the $k$-mer set into strand specific clusters, we computed the reverse complement of one cluster and merged the result with the other one, and thereby generated a final set of positive $k$-mers that are more likely to be located on the same strand. Next, we performed a progressive multiple sequence alignment with the NUC44 scoring matrix. We removed the trailing aligned positions if they were supported by less than 15% of the elements in the final $k$-mer set. Once the alignment is done, we need to generate informative features for our downstream classifier. To do that, we first build a PWM by computing the nucleotides' relative frequencies at each position. Position weight matrices, as described, only give a simplistic characterization of the motifs and hence are not well suited for prediction of the DNA binding site. Therefore, models with higher computational capacities are needed that can capture the dynamics of the binding sites for each transcription factor. On the other hand, as reported in literature [5], there is no linear relation between the binding affinity and the semi-quantitative readouts of microarray intensities. This makes the robust prediction of specificity or affinity a challenging task. To address this challenge, we used support vector machines (SVMs) to discriminate between probes with high measured affinity and others which are not as highly bound. In fact, we selected the top 1000 probes and labeled them as positive and the next 5000 probes and assigned them to the negative class. Note that for the negative class we selected the probes with highest scores after positive ones. By doing that we are indirectly instructing the SVM solver to pick the support vectors (SVs) that are helpful in recognizing those subtle differences that make a sequence favored slightly over other competing ones. Once the model is trained, we can treat the prediction scores as a measure of binding affinity between each DNA sequence and the TF. We used the derived PWM as a seed to generate features for our classifier. Specifically, using the PWM, we scanned probes in the training set and recorded the position which resulted in the highest score across the probe. Then, we adopted a one-hot coding scheme to represent each base in the selected subsequence. This serves as a binary mask for the PWM matrix through which we can efficiently encode both the PWM features and the positional information into the feature vector that is a flattened version of the masked matrix. Once the feature matrix is built, we feed it into our model which exploits a polynomial kernel for capturing non-linear dependencies between each base position. Lastly, due to the non-deterministic outcome of the alignment step, we repeat the algorithm steps multiple times and store the model which yields the highest performance (see the next section) on the training set.

III. RESULTS

To make consistent comparisons across models, we retrieved the same PBM experiment data (i.e., Rap1, Oct-1, CEH-22, Cbf1 and Zif268) from the UniPROBE database that were previously used in the baseline studies. For each TF, UniPROBE contains two sets of experiments which are named array #1 and #2. Each array in the database represents the outcome of an independent assay with a different set of synthetically generated probes. To assess our pipeline, we trained our model as well as the baselines, once for each array and tested their performance on the other one. For

Figure 2: Predicted ranks of the top 100 positive probes (black lines) in arrays #1 and #2 of Cbf1; RankMotif++ (first column), KmerHMM (second column) and MotifMark (third column).

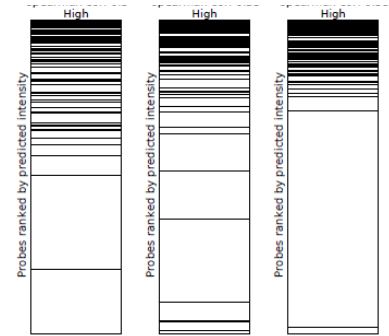

(a) Prediction on array #2 (trained on array #1)

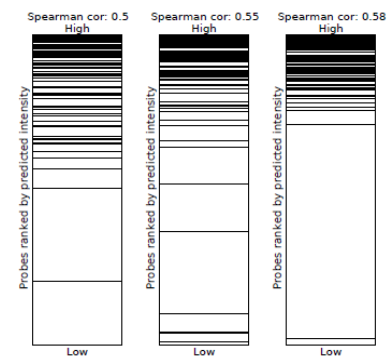

(b) Prediction on array #1 (trained on array #2)

RankMotif++, the motif length range and the number of repeats were set to 9-12 and 5, respectively. Moreover, we found out (data not shown) that for both KmerHMM and MotifMark, $k = 8$ is the best length for the seed $k$-mers that we align in the first stage of these pipelines. The rest of parameters were set according to the best practices suggested by these studies. To make a fair benchmarking of the proposed pipeline, we assessed the performances of our model using the same criteria used in the baseline studies as described below. In order to compile a universal test set, we adopted the same strategy as in [5, 6] to select positive probes. Note that each of the baseline methods follow a specific protocol for compiling a training set. For MotifMark, we simply select the top 1000 probes as positive probes and the next 5000 as negative ones and set $|M| = 200$.

To assess the generalization capability of each method we computed the Spearman rank correlation between the measured probe intensities and the score predicted for each probe in the test array using the trained models. Table 1 presents the resulting scores for RankMotif++, KmerHMM, PWM model learned by MotifMark and finally, the complete pipeline. According to the table, MotifMark is either the best performer, sometimes by a large margin (e.g., when predicting on the second array of Rap1), or the runner-up following the best model closely. The improvement attained by the MotifMark over its underlying learned PWM is a proof of concept for the existence of a non-linear relationship between base-positions in motifs. An exception to this rule is the second array of Rap1, which can be attributed to the overfitting that is taken place due to a very small training set of only 44 positive probes.

Calculating the correlation between predicted scores and normalized intensities is not a direct way to visualize model performance. In fact, sometimes what is more important is to be able to predict the top performing $k$-mers from a comprehensive pool of DNA probes. In light of that, for each test array we marked the top 100 probes based on their measured intensities and computed their predicted rank after applying each method to the whole array.

TABLE 1: SPEARMAN RANK CORRELATION COEFFICIENTS: EVALUATION OF DIFFERENT METHOD ON ARRAYS #1 AND #2

| Dataset | Test Array | Spearman Correlation | | | |
|---|---|---|---|---|---|
| | | RnkMtf | K-HMM | $PWM^{MM}$ | MotifMark |
| Cbf1 | #1 | 0.50 | 0.56 | 0.52 | **0.57** |
| | #2 | 0.50 | 0.55 | 0.44 | **0.59** |
| CEH-22 | #1 | 0.40 | 0.41 | 0.46 | **0.48** |
| | #2 | 0.40 | 0.28 | 0.28 | **0.44** |
| Oct-1 | #1 | 0.27 | **0.28** | 0.20 | 0.27 |
| | #2 | 0.26 | **0.37** | 0.28 | 0.33 |
| Rap1 | #1 | 0.25 | 0.24 | 0.24 | **0.31** |
| | #2 | 0.29 | 0.24 | **0.54** | 0.50 |
| Zif268 | #1 | 0.31 | 0.31 | 0.39 | **0.42** |
| | #2 | 0.29 | **0.36** | 0.32 | 0.34 |

In the ideal case we would like to have the predicted ranks placed as the top 100 samples. Figure 2 visualizes the described criterion for Cbf1 protein, using bar charts. Each line in the chart represents one of the top 100 probes, with its height representing the predicted score. In other words, the more the lines are focused towards the top of chart the better the predictions would be. As implied by this figure, the binding preference that is predicted by MotifMark is more concentrated towards the top of the chart compared to the other methods. A noteworthy point is that, despite the fact that the correlation scores across each method are not largely different, MotifMark's predictions turns out to be more favorable.

IV. CONCLUSION

In this article we propose a new approach for predicting DNA binding affinity of proteins to the DNA probes using principles from graph theory and machine learning. Through assessments and comparisons we show that the new design beats the most accurate classical prediction tools that we are aware of, sometimes by a large margin. The proposed method also runs approximately an order of magnitude faster than the baseline methods selected for our comparisons (data not shown). In our view, the promise of this work is mostly due to 1) using graph theory to improve the alignment of $k$-mers though the heuristics that we designed to predict the strand associated with each $k$-mer and 2) exploiting a non-linear kernel to capture the complex relation between motif base positions while at the same time benefiting from the information that is available in the inferred PWM.

Despite its promise, MotifMark does not take into account the contributions made by individual $k$-mers in each probe. Moreover, we did not model the impact of $k$-mers' locations inside each probe. Notwithstanding the argument made in [6], our internal analyses suggest that the location of the best performing $k$-mer in a probe conveys useful information. Therefore, as a future work one can incorporate these facts into an improved model to boost the prediction performance.